\documentstyle[eqsecnum,aps,epsf]{revtex}

\begin{document}
\draft
%\input epsf
%\twocolumn
\title{SPIN EFFECTS IN GRAVITATIONAL RADIATION
BACKREACTION\hskip5cm I. THE LENSE-THIRRING APPROXIMATION}
 
\author{L\'aszl\'o \'A. Gergely, Zolt\'an I. Perj\'es and M\'aty\'as
Vas\'uth} \address
{KFKI Research Institute for Particle and Nuclear
Physics, Budapest 114, P.O.Box 49, H-1525 Hungary}
\maketitle
\begin{abstract}
   The gravitational radiation backreaction effects are considered in
the Lense-Thirring approximation. New methods for parameterizing the
orbit and for averaging the instantaneous radiative losses are developed.
To first order in the spin $S$ of the black hole,
both in the absence and in the presence of gravitational radiation,
a complete description of the test-particle orbit is
given. This is achieved by two improvements over the existing
descriptions. First, by introducing new angle variables with a
straightforward geometrical meaning. Second, by finding a new parametrization
of a generic orbit, which assures that the integration over a radial
period can be done in an especially simple way, by applying the residue
theorem.
The instantaneous gravitational radiation losses of the system are
computed using the formulation of {\em Blanchet, Damour and Iyer(1989)}.
All losses are given both in terms of the dynamical constants of motion
and the properly defined orbital elements $a,e,\iota$ and $\Psi_0$.
The radiative losses of the constants characterizing the Lense-Thirring
motion, when suitably converted, are in agreement with earlier
results of {\em Kidder, Will and Wiseman(1993), Ryan(1996)
and Shibata(1994)}. In addition, the radiative losses of two slowly
changing orbital elements $\Psi_0,\Phi_0$ are given in order to
complete the characterization of the orbit.
 
\end{abstract}
\pacs{PACS Numbers:04.25.Nx, 04.30.Db }

\section{Introduction}
   The determination of the signal template of the radiating
companion of a black hole is of great importance for
the forthcoming interferometric gravitational observatories.
An inspiralling neutron star or a several
solar-mass black hole orbiting a massive black hole, as
well as a particle constituent contributing to the radiation of the debris
trapped by the black hole are notable examples of these companions.
Either is the case, one wants to describe the evolution of the
orbit of a test particle in the neighborhood of a black hole under
the influence of gravitational radiation backreaction. In a highly
idealized picture, the orbit
of the particle is a Carter geodesic\cite{Carter} characterized
by four constants of the motion, the energy, the total angular
momentum, the rest-mass of the particle and the separation
constant. In the spirit of perturbation theory, the radiation
backreaction effects may be taken into account by evolving these
constants of the motion, thus picturing the trajectory of the
particle by a sequence of geodesic orbits.
 
   The history of the orbit may be divided in separate epoches.
For example, the first thing that happens to a particle trapped
by a nonrotating {\em Schwarzschild} black hole on an arbitrary
but distant orbit is that it loses eccentricity due to radiation losses,
and the radius
shrinks until the close region of the last stable circular orbit
is approached. There one has to replace the adiabatic picture with
one valid in the strong field regime.
 
  The description of a generic orbit in the strong field region of a
spinning {\em Kerr} black hole is difficult
(as has been pointed out by Thorne\cite{Thorne})
due to the special nature
of the separation constant which lacks a straightforward geometrical
interpretation. Recently, progress in treating the separation constant
has been reported by Ori\cite{Ori}.
 
   There is a considerable ease in the description of
special orbits such as equatorial or quasicircular, and in fact,
several papers have obtained the signal templates for such
special orbits. In order that any of these special
orbits can be credited as significant contributor to the gravitational
signal, one has to make sure that the Carter constant evolves to
the envisaged special values.
 
  Several authors have recently published results on radiative
losses by systems of spinning masses.
Kidder, Will and Wiseman\cite{KWW} formulate the problem by using
the Lagrangian (proposed by Barker and O'Connell\cite{BOC}) for
bodies with finite masses and spins. They and Kidder \cite{Ki}
compute the spin-orbit and spin-spin contributions to the
momentary radiative loss of energy, linear and angular momenta.
In the test particle limit their results have subsequently been
used by Shibata\cite{Sh}, in his study of equatorial orbits, and
by Ryan\cite{Ry,Ry2}. By computing the backreaction forces, Ryan obtains
the power and angular momentum losses for circular and generic orbits.
 
   In our work, we have taken up the evolution of the orbit in the
far field region of the black hole. The trajectory of the
idealized, nonradiating particle is described by the
Lense-Thirring picture\cite{LT,LL}. This picture emerges in the test
particle limit of the scenario described in \cite{KWW}. The spin ${\bf
S}$ of the black hole is fixed to the $z$ axis.
The orbit is viewed approximately as an ellipse precessing in a plane
which rotates about ${\bf S}$. The motion is characterized by three
constants: the energy $E$, the magnitude $L$ and spin projection $L_z$
of the orbital angular momentum, and by two slowly changing angle
parameters $\Psi_0$ and $\Phi_0$. The Lense-Thirring
approximation is linear in the spin terms, thus the (additive)
Schwarzschild effects may be neglected in this treatment.
 
  In an earlier paper\cite{GP}, we have computed the radiation losses
from the mass quadrupole tensor in the Lense-Thirring approximation,
by the orbit smoothing method of Landau and Lifshitz\cite{LL}.
However, one may be rightly concerned about the degradation of accuracy
of this computation, as a price of the relative simplicity
introduced by the smoothing method.
 
  In the present paper, we drop the orbit smoothing, and
we include the velocity quadrupole tensor $J_{ik}$ in the
description of the radiative losses. In the test particle limit,
this (formally) corresponds to the 1PN approximation\cite{Ry}.
The rest of the paper is organized as follows. In Sec. 2 we
describe the orbital motion in the Lense-Thirring approximation.
In terms of a set of new angular variables (related to Euler rotations),
the equations of motion take a particularly simple form: all time
derivatives depend on the radial variable $r$ alone.
 
 In Sec. 3, we explore the advantages of having a pure radial equation
of motion. We find the turning points at $\dot r=0$ which allow us to
parameterize the radial motion by trigonometric functions.
Integrations over the time can be replaced by integrations over
the parameter, which are straightforward in principle, but can be
messy in practice. We evade such complications by finding suitable
parametrizations. The first parameter $\xi$ we introduce is a
generalization of the eccentric anomaly of the Keplerian orbits.
It is employed for computing the half period defined as the time spent
by the particle between consecutive turning points. For the purpose
of averaging the radiative losses, and for determining the change in the
new angular variables, a generalized true anomaly parameter $\chi$ is
properly defined. In terms of this parameter, when integration by the residue
theorem is performed, the only pole of the integrands will be in the
origin.
 
In Sec. 4, we compute the rates of change of the constants of motion
due to radiation backreaction, to first order in ${\bf S}$, by
employing the radiative multipole tensors of Kidder, Will and
Wiseman\cite{KWW}, which originate in the Blanchet-Damour-Iyer
formalism\cite{BD,DI}.
As we are interested in the cumulative effects of the secular motion,
we compute averaged losses over the period of radial motion.
We give the losses in an unambiguous manner in terms of the constants of
motion $E,L$ and $L_z$.
 
 As has been pointed out by Ryan \cite{Ry2}, the definition of the
orbit parameters $a$ and $e$ is subject to ambiguities. In Sec. 5,
we define the semimajor axis and eccentricity. With advantages
explained in the main text, our definitions differ both from Ryan's
and the ones used in the Lense-Thirring paper. We rewrite all losses
in terms of our orbital parameters $\iota,a$ and $e$.
 
  In the concluding remarks, we complete the
characterization of the changes caused by gravitational radiation
by giving the change in the $\Psi_0,\Phi_0$ angle parameters
over one orbit. We then compare our results
with those of other authors.
 
  We wish to make some comments on the expansion procedure we have
adopted. Earlier works employ either the expansion
parameter $\epsilon\approx v^2\approx {m/ r}$ or the
inverse ${1/ c}$ of the speed of light.
Since we eliminate $v^2$ and $\dot r^2$ by
use of the first integrals of the motion,
we choose to retain $G$ and $c$ for bookkeeping of the orders in our
formalism. The correction terms in the Lagrangian, in the
equations of motion and radiation losses will carry
an extra $1/c^2$ factor. In our computation,
we keep only the terms containing the spin $S$ from among these
correction terms, thus we may as well choose $S$ as our
expansion parameter.
 
  Our treatment of radiation backreaction effects can be generalized
beyond the Lense-Thirring approximation. In a follow-up paper\cite{GPV}
we will apply our averaging method and parameterization for computing
the backreaction effects on a bound system of two finite masses.
 
\section{The Orbit in the Lense-Thirring approximation}
 
  We consider the orbit of a small particle of mass $\mu$ about a
body with mass $M$ and angular momentum vector $\bf S$. In the
Lense-Thirring approximation\cite{LT}, the Lagrangian of the
system has the form\cite{LL}
\begin{equation}
{\cal L}=\frac{\mu{\bf \dot r^2}}{2}+{G \mu M\over r}+\delta{\cal L} \ ,
\end{equation}
where
$r=\vert{\bf r}\vert$, a dot denotes $d/dt$ and
the term describing (perturbative) rotation effects is
\begin{equation}
\label{eq:delag}
\delta {\cal L}={2G\mu\over c^2r^3}{\bf S}\cdot({\bf\dot r}\times{\bf r}) \ .
\label{delag}\end{equation}
The Cartesian coordinates of the test particle are
${\bf r}=\{x,y,z\}$, with origin chosen
at the black hole.
 
  Hence the orbital momentum
\begin{equation}
\label{eq:L}
{\bf L}={\bf r}\times{\bf p}\ \label{l}
\end{equation}
and Runge-Lenz vector
\begin{equation}
\label{eq:A}
{\bf A}={{\bf p}\over \mu}\times{\bf L}-{G\mu M{\bf r}\over r}\
 \label{a} \end{equation}
satisfy the equations of motion, respectively
\begin{equation}
\label{eq:Ldot}
{\bf\dot L}={2G\over c^2r^3}{\bf S}\times{\bf L}\label{ldot}
\end{equation}
\begin{equation}
\label{eq:Adot}
{\bf\dot A}={2G\over c^2r^3}{\bf S}\times{\bf A}
           +{6G\over c^2r^5\mu}({\bf S\cdot L)\ r}\times{\bf L}\
.\label{adot} \end{equation}
Here
${\bf p}={\partial{\cal L}/\partial{\bf\dot r}}$
is the momentum of the orbiting mass. Note that the
angular momentum ${\bf L}$ undergoes a pure rotation about the
spin with angular velocity:
\begin{equation}
\Omega_S={2G\over c^2r^3} S \ .
\end{equation}
 
From (\ref{l}) and (\ref{a}), the vector ${\bf r}$ can be written in
terms of the basis $\{{\bf A}\ ,{\bf L}\times{\bf A}\}$ in the
plane with normal ${\bf L}$:
\begin{equation}
{\bf r}={L^2-G \mu^2 M r\over \mu A^2}{\bf A}+
      {{\bf r}\cdot{\bf p}\over \mu A^2} {\bf L}\times{\bf A} \ ,
\end{equation}
where $A=\vert{\bf A}\vert$. Inserting this expression of ${\bf
r}$ in (\ref{adot}), the direction ${\bf A}/A$ of the Runge-Lenz
vector is
found to rotate about a linear combination of the vectors ${\bf S}$ and
${\bf L}$:
\begin{equation}
\label{eq:averst}
\left({{\bf \skew{39}{\dot}{A}}\over
A}\right)=\left[{2G\over c^2r^3}{\bf S}
     -{6G(L^2-G \mu^2 M r)\over c^2\mu^2A^2r^5}({\bf S\cdot L})
{\bf L} \right]\times{{\bf A}\over A}
\ .\label{averst} \end{equation}
Thus the direction vector rotates about ${\bf S}$ with
the angular velocity $\Omega_S$ (as
does ${\bf L}$) and about ${\bf L}$ with the angular velocity:
\begin{equation}
\label{omegaL}
\Omega_L=-{6GL(L^2-G \mu^2 M r)\over c^2\mu^2A^2r^5}({\bf S\cdot
 L}) \ .
\label{omegal}\end{equation}
 
Although the orbital momentum and the Runge-Lenz vector are not
conserved (unlike in the zeroth order in ${\bf S}$), here also
there are constants of motion.
Because of the fact that the potential is axisymmetric and
stationary, the orbital momentum component $L_z$ and the energy
$E={\bf \dot r}{\partial{\cal L}/\partial{\bf\dot r}}-{\cal L}$
are conserved. Since the motion of ${\bf L}$ is a pure
rotation, its magnitude $L$ is also constant:
\begin{equation}
\dot E=\dot L_z =\dot L=0\ .
\end{equation}
 The constants of the motion are
\begin{eqnarray}
\label{eq:E}
E&=&{\frac{\mu}{2}[\dot
r^2+r^2(\dot\theta^2+\sin^2\theta\ \dot\varphi^2)]}-{G \mu M\over
r}\label{e}\\
\label{eq:Lz}
L_z&=&\mu r^2\sin^2\theta\
\dot\varphi-2\frac{G\mu}{c^2r}S\sin^2\theta \label{Lz} \\
\label{eq:L2}
L^2&=&{\mu^2}r^4
(\dot\theta^2+\sin^2\theta\
\dot\varphi^2)-4{G\mu^2S\over c^2}r\sin^2\theta\ \dot\varphi  \ .
\label{L2} \
\end{eqnarray}
(Here we are using polar coordinates,
$x=r\sin\theta\cos\varphi$, $y=r\sin\theta\sin\varphi$,
$z=r\cos\theta$).
As a consequence, the angle $\iota$ subtended by ${\bf L}$ and
${\bf S}$ is also conserved:
\begin{equation}
\label{eq:iota}
\cos\iota={L_z\over L}\ .
\label{iota}
\end{equation}
The first integrals of the equations of motion follow from
(\ref{eq:E})-(\ref{eq:L2}) by simple algebra (we need not
recourse to the Hamilton-Jacobi formalism).
Thus the equation of the radial motion takes the first-order form,
decoupled from the angular degrees of freedom:
\begin{equation}
\label{eq:radial}
\dot r^2=-\frac{L^2}{\mu^2
r^2}+2\frac{GM}{r}+2\frac{E}{\mu}-4\frac{GL_z}{c^2\mu r^3}S \ ,
\label{radial}
\end{equation}
the equations for the angular variables $\theta,\varphi$, however,
are coupled among themselves and with the $r$ motion:
\begin{equation}
\label{eq:polar}
\dot\theta^2={L^2\over\mu^2r^4}\left(1-{\cos^2\iota\over
\sin^2\theta}\right)\ ,\qquad
\dot\varphi={L_z\over\mu r^2\sin^2\theta}+{2GS\over c^2r^3} \ .
\label{polar}
\end{equation}
 
  There is an alternative description of the motion in the
Lense-Thirring approximation which bears a more intimate
geometrical relation to the picture provided by perturbation
theory. We shall introduce this formalism by performing
time-dependent Euler rotations about the origin so as to place
the Kepler ellipse in its momentary orientation:
\begin{equation}
{\bf r}=R_z(\Phi)R_x(\iota)R_z(\Psi){\bf r_0}\ .
\end{equation}
Here
\begin{equation}
\label{eq:newvar}
{\bf r_0}=\left(\matrix{r \cr 0 \cr 0 }\right)
\label{newvar}
\end{equation}
is the initial position of the particle on the plane perpendicular to
${\bf L}$\ (although ${\bf L}$ is different from ${\bf L}_N=\mu
{\bf r}\times{\bf v}$, the condition ${\bf r} \perp {\bf L}$ is
fulfilled). A rotation about the Cartesian $x$ and $z$ axes is
given, respectively, by
 
\begin{equation}
R_x(\iota)=\left(\matrix{1  &  0          & 0            \cr
              0  & \cos\iota & -\sin\iota \cr
              0  & \sin\iota &  \cos\iota  }\right)\ ,
\qquad
R_z(\Phi)=
\left(\matrix{\cos\Phi & -\sin\Phi & 0 \cr
              \sin\Phi &  \cos\Phi & 0 \cr
              0          &  0          & 1 }\right)
\ .\end{equation}
 
The coordinates of the particle are written in terms of the Euler
angles $\Phi,\iota$ and $\Psi$ as
\begin{eqnarray}
\label{eq:newvar2}
x&=&r(\cos\Phi\cos\Psi-\cos\iota\sin\Phi\sin\Psi) \nonumber\\
y&=&r(\sin\Phi\cos\Psi+\cos\iota\cos\Phi\sin\Psi) \nonumber\\
z&=&r \sin\iota\sin\Psi \ .
\label{newvar2}
\end{eqnarray}
 
Hence we get the relations
 
\begin{equation}
{\dot\theta}=-{\sin\iota\over
\sin\theta}\cos\Psi{\dot\Psi}\ ,\qquad
\dot\varphi=\dot\Phi+{\cos\iota\over
\sin^2\theta}\dot\Psi\ ,\qquad
\sin^2\theta=1-\sin^2\iota \sin^2\Psi
\ ,\end{equation}
 
\noindent
which allow us to express the time derivatives of the new
variables:
 
\begin{eqnarray}
\label{eq:psit}
\dot\Psi&=&{\frac{L}{\mu r^2}} \label{psit}\\
\label{eq:phit}
\dot\Phi&=&{\frac{2GS}{c^2 r^3}} \label{phit}
\ .\end{eqnarray}
Note that $\dot\Phi=\Omega_S$ and $\dot\Psi$ carries the
interpretation of angular velocity of rotation about ${\bf S}$
and in the plane perpendicular to ${\bf L}$ respectively.
An important feature of the new coordinates ${\Psi,\Phi}$ is
that their derivatives appear linearly in the equations
of motion (\ref{psit}) and (\ref{phit}). This is not the
case with the polar coordinates (\ref{polar}) because the
root of ${\dot \theta}^2$ cannot be extracted. The geometrical
interpretation of the angle variable ${\Psi}$ assures that it is
monotonously changing. We choose $\Psi$ to increase
with the motion evolving.
 
  There is a freedom in the precise way one introduces the three
Euler-angle variables in place of the polar angles.
Even though the equations of motion are made simple by use of
the angular variables (\ref{newvar}-\ref{newvar2}), the description of
the momentary plane of orbit is tied to a different choice of the
Euler angles $(\Psi^\prime,\iota^\prime,\Phi^\prime)$,
where the initial $z$ axis is chosen along ${\bf L}_N$ rather than
along ${\bf L}$. These angular variables appear (with the appropriate
change in notation) in the Lense-Thirring paper.
The angle $\iota^\prime$ is not a constant, and the resulting
equations are less simple.
The primed angles will not be employed in our computations, and we
relegate their discussion to the Appendix.
 
\section{Parametrization of the orbit}
 
  We now proceed with the solution of the equations of motion.
First consider the radial equation (\ref{radial}).
For the turning points $\dot r=0$, we get the cubic equation
\begin{equation}
\label{eq:cubic}
2E\mu r^3+2GM\mu^2r^2-L^2r-4{G\over c^2}\mu L_zS=0\ .
\label{cubic}\end{equation}
 In the no-spin limit, $S=0$, the solutions are (the
unphysical) $r=0$ and
\begin{equation}
\label{eq:rpm}
r_{\pm}=\frac{-GM\mu\mp A_0}{2E}\ ,
\label{rpm} \end{equation}
where $A_0$ is the length of the Runge-Lenz
vector to zeroth order:
\begin{equation}
\label{eq:A0}
A_0=\sqrt{G^2M^2\mu^2+{2EL^2\over\mu}} \ .
\label{a0}\end{equation}
 
 When a small spin term is present,
the roots of (\ref{cubic}) will have a slightly
different form from (\ref{rpm}):
$0+\epsilon$ and $r_{\pm}+\epsilon_{\pm}$, where $\epsilon$ and
$\epsilon_{\pm}$ are assumed to be small. Note that the constants
$E,L$ and $L_z$ will not coincide with their zeroth order limits.
The constant $A_0$ is still defined by (\ref{a0}), although it
is not identical with the length $A$ of the
Runge-Lenz vector (which is not constant). Writing the cubic equation as
$2E\mu(r-r_+-\epsilon_+)(r-r_--\epsilon_-) (r-\epsilon)=0$,
we get
 
\begin{equation}
2E\mu[r^2-r(r_++r_-)+r_+r_-]r+2E\mu[-r^2(\epsilon_++\epsilon_-
+\epsilon)+r(\epsilon_+r_- +\epsilon_-r_++\epsilon(r_++r_-))
-\epsilon r_+r_-]=0\ .
\end{equation}
The first term is the zeroth-order part of the original equation,
written in terms of the roots, thus it equals $(4G/c^2)\mu
L_zS$. Inserting the chosen form of the roots we get
\begin{equation}
\epsilon=-\frac{4G\mu L_zS}{c^2L^2},\qquad
\epsilon_{\pm}=\frac{2G\mu L_zS}{c^2L^2}\Bigl(
1\mp\frac{GM\mu}{A_0}\Bigr)\ .
\end{equation}
The orbit is supposed to lie at large distance, thus
the first root will not occur.
 
  Using the above results, a parameter $\xi$ generalizing
the eccentric anomaly of the Kepler orbits is readily introduced:
\begin{equation}
\label{eq:xi}
r=-\frac{G\mu M}{2 E}+\frac{2G\mu L_zS}{c^2L^2}
+\Biggl(\frac{A_0}{2 E}
   +\frac{2G^2\mu^2 ML_zS}{c^2A_0L^2}
 \Biggr)\cos\xi
\label{xi}\end{equation}
such that $r_{{}^{max}_{min}}$ is at $\xi=\pi$ and $\xi=0$,
respectively. From here:
\begin{equation}
{d r\over d\xi}=-\Biggl(\frac{A_0}{2
E}+\frac{2G^2\mu^2ML_zS}{c^2A_0L^2}
 \Biggr)\sin\xi \ .
\end{equation}
For equatorial orbits, this parametrization has been employed by
Shibata\cite{Sh}.
Expressing ${1/\dot r}$ from (\ref{radial}) and linearizing
in $S$, the expression for $dt/d\xi$ is found:
\begin{equation}
\label{eq:txi}
{dt\over d\xi}={1\over\dot r}{dr\over d\xi}=
{\mu^2(GM\mu-A_0\cos\xi)\over (-2\mu E)^{3\over 2}} +
{2G^2M\mu^3L_zS\cos\xi\over (-2\mu E)^{1\over 2}L^2A_0c^2} \ .
\label{txi}\end{equation}
 
Integration of (\ref{txi}) from 0 to $2\pi$ gives the orbital
period:
\begin{equation}
\label{eq:period}
T=2\pi{GM\mu^3\over (-2\mu E)^{3\over 2}} \ .
\label{period}\end{equation}
 
However the parametrization (\ref{eq:xi})
is inconvenient for carrying out the averaging of the losses,
because it yields a complicated array of poles of the integrands.
Hence in the sequel we will, in place of
(\ref{xi}), find a parametrization
$r=r(\chi)$ satisfying the following two criteria:
\begin{eqnarray}
(a)&{}& \quad r(0)=r_{min}\quad and\quad r(\pi)=r_{max}\\
(b)&{}& \quad \frac{dr}{d(\cos\chi)}=-(\gamma_0+S\gamma_1)r^2 \ ,
\end{eqnarray}
where $\gamma_0,\gamma_1$ are constants.
That is, we want to keep the properties of the true anomaly
parametrization of the Kepler orbit.
Property (b) generalizes Kepler's second law for the
area\cite{Ptolemy} .
The unique parametrization satisfying both (a) and (b)
is:
 
\begin{equation}
\label{eq:chi}
r= \frac{L^2}{\mu (GM\mu+A_0\cos\chi)}+\frac{4GL_zS}{A_0L^2c^2}\,
\frac{A_0(2G^2M^2\mu^3+EL^2)+GM\mu(2G^2M^2\mu^3+3EL^2)\cos\chi}
{(GM\mu+A_0\cos\chi)^2} \ .
\label{chi}\end{equation}
Hence
\begin{equation}
{d r\over d\chi}=\Biggl[\frac{\mu A_0}{L^2}-
\frac{4G^2M\mu^3L_zS}{A_0L^6c^2}(2G^2M^2\mu^3+3EL^2)
 \Biggr]r^2\sin\chi
\end{equation}
and to first order in $S$:
\begin{equation}
\label{eq:tchi}
{dt\over d\chi}={1\over\dot r}{dr\over d\chi}=
{\mu r^2\over L}
\Bigl[1-{2G\mu^2 L_zS\over c^2L^4}(3GM\mu+A_0\cos\chi)\Bigr] \ .
\label{tchi}
\end{equation}
 
We will need also the relation of the type $\Psi=\Psi(\chi)$. This
can be obtained from integration of (\ref{psit}):
\begin{equation}
\label{eq:psi}
\Psi=\Psi_0+\chi-\frac{2G\mu^2L_zS}{c^2L^4}
                    (3GM\mu\chi+A_0\sin\chi) \ .
\label{psi}\end{equation}
 
  The constant of integration $\Psi_0$ measures the angle subtended by the
semiminor axis ($\chi=0$) and the node line at $\Psi=0$. In a perturbative
picture, the orientation of the momentary ellipse of the orbit is
reinterpreted at each revolution, and the value of the
integration constant is subject to a corresponding shift.
 
Our parametrization (\ref{chi}) makes especially simple the
integration over one period of all expressions $F=F(r,\Psi)$
containing in the denominator no $\Psi$ dependence and no other
$r$ dependence than $r^{2+n}$, with $n$ a positive integer.
Note that (\ref{omegal}), (\ref{psit}) and (\ref{phit}) have this
property. Further, as will be shown in the next section,
the instantaneous radiative losses of the constants of motion
share this property. In all of these cases one has:
\begin{equation}
\int_0^{T}F(r(t),\Psi (t))dt=
\int_0^{2\pi}F(r(\chi),\Psi(\chi)){dt\over d\chi}\ d\chi \ ,
\end{equation}
where the $r$ dependence of the denominator of the integrand is
especially simple: $r^n$. The {\em average} over one period is
\begin{equation}
<F>=\frac{1}{T}\int_0^{T}F(r(t),\Psi (t))dt \ ,
\end{equation}
with $T$ given by Eq. (\ref{period}).
 
The integrals may conveniently be evaluated by use of the residue
theorem. One introduces the complex variable $\zeta=e^{i\chi}$,
and the integral is taken over the unit circle in the $\zeta$ plane.
The value of the integral is given by the sum of the residues of
the poles inside the circle. In all relevant cases our
parametrization assures that the only pole is at $\zeta=0$.
 
Simple examples are the integrations of (\ref{phit}), (\ref{psit})
and (\ref{omegal}):
\begin{eqnarray}
\label{eq:dephi}
\Delta\Phi&=&\qquad\  2\pi S\frac{2G^2M\mu^3}{c^2L^3}
\label{dephi} \\
\label{eq:depsi}
\Delta\Psi&=&2\pi\ -\ 2\pi S\frac{6G^2M\mu^3L_z}{c^2L^4}
\label{depsi}   \\
\label{eq:shift}
\Delta\Omega_L&=
            &\qquad -\ 2\pi S\frac{6G^2M\mu^3L_z}{c^2L^4}\ .
\end{eqnarray}
   Here (\ref{dephi}) is
the precession angle of the orbital angular momentum ${\bf
L}$ about ${\bf S}$.
The interpretation of the variable $\Psi$ as the polar angle in the
plane perpendicular to ${\bf L}$ is in accordance with
the identical first order terms in (\ref{eq:depsi}) and (\ref{eq:shift}).
The ellipse is found to counterrotate (for $L_z>0)$. The
periastron shift is a combination of (\ref{eq:dephi}) and
(\ref{eq:shift}). Both $\Delta\Psi$ and $\Delta\Phi$
will be unchanged in the averaged-motion approximation
of Landau and Lifsitz \cite{LL,GP}.
 
    Our angles are related to the polar angle $\varphi$ for equatorial
orbits ($\iota=0$) by $\varphi=\Phi+\Psi$. This relation helps
comparison with Shibata's \cite{Sh} expression (2.19) for the
periastron shift $\Delta\varphi$.
 
  Ryan\cite{Ry2} has introduced a
parametrization $r=r(\psi_R)$ which,
when rewritten in terms of the constants of motion
$E,L$ and $L_z$ (employing his definitions for the orbital parameters),
will read:
 
\begin{equation}
\label{eq:psiryan}
r= \frac{L^2}{\mu (GM\mu+A_0\cos\psi_R)}+\frac{4GL_zS}{A_0L^2c^2}\,
\frac{A_0(2G^2M^2\mu^3+EL^2)+GM\mu(2G^2M^2\mu^3+3EL^2)\cos\psi_R
-(\mu A_0^3/2)\sin^2\psi_R}
{(GM\mu+A_0\cos\psi_R)^2}  \ .
\label{psiryan}\end{equation}
 
This differs from our parametrization (\ref{chi}) by the last
$sin^2\psi_R$ term, which shows that Ryan's parameter $\psi_R$ and
our $\chi$ reach the turning points simultaneously. In fact both
(\ref{xi}) and (\ref{chi}) can be supplemented by arbitrary terms
involving a $sin$ squared factor, and they still will satisfy (a).
However none of these parameters, and among those Ryan's, will
satisfy the generalized second Kepler law (b).
 
\section{Instantaneous and Averaged Losses}
 
   The mass and current quadrupole tensors
of a test particle with coordinates $r_i=(x,y,z)$ are
\begin{equation}
I_{ij}=\mu (r_i r_j)^{STF}\
\end{equation}
\begin{equation}
J_{ij}= - \mu \bigl[r_i ({\bf r}\times {d{\bf r}\over
dt})_j\bigr]^{STF} +
       {\frac {3 \mu}{2 M}} (r_i S_j)^{STF} \ ,
\end{equation}
where STF means symmetrization and trace removal in the free indices.
The rates of the energy and the angular momentum losses in
the test particle 1PN approximation are, respectively\cite{KWW},
 
\begin{equation}
{dE\over dt}= - {G\over 5c^5}\Bigl({d^3I_{ij}\over dt^3}
                            {d^3I_{ij}\over dt^3}
                      +{16\over 45c^2}{d^3J_{ij}\over dt^3}
                            {d^3J_{ij}\over dt^3}\Bigr)\label{q}
\end{equation}
\begin{equation}
{dL_i\over dt}=-{G\over c^5} \epsilon_{ijk}
       \Bigl({2\over 5} {d^2I_{jr}\over dt^2}
                            {d^3I_{kr}\over dt^3} +
        {32\over 45c^2}{d^2J_{jr}\over dt^2}{d^3J_{kr}\over dt^3}
                            \Bigr)\label{Li} \ .
\end{equation}
  Evaluating the right-hand sides by use of
(\ref{radial}), (\ref{psit}) and (\ref{phit})
and dropping higher-order terms in $S$, the total power is
\begin{eqnarray}
{dE\over dt} =
&-& {{8\,G^{3}M^{2}\over 15\,c^{5}r^{6}}\left (2\,E\mu\,r^{2}+2\,GM\mu^{2}r+11\,L^{2}
\right )}+{{8\,SG^{3}ML\cos\iota\over 15\,\mu\,c^{7}r^{8}}\left (
20\,E\mu\,r^{2}-12\,GM\mu^{2}r+27\,L^{2}\right )} \ .
%\nonumber\\
%&-& {{8\,G^{3}M^{2}L^{2}\over 45\,\mu^{2}c^{7}r^{8}}\left
%(8\,E\mu\,r^{2}+8\,GM\mu^{2}r-3\,L^{2} \right )} .
\end{eqnarray}
 
This is independent of the angles $\Psi$ and $\Phi$.
The loss in $L$ is evaluated by
$2L\delta L=\delta L^2=\delta(L_iL_i)=2L_i\delta L_i$
and it has the form
\begin{eqnarray}
{dL\over dt} =
&+& {{8\,G^{2}LM\over 5\,\mu\,c^{5}r^{5}}\left (2\,E\mu\,r^{2}-3\,L^{2}\right )}
+{{8\,G^{2}S\cos\iota\over 15\,c^{7}\mu\,r^{6}}\left (12\,GEM\mu^{2}r^{2}
+3\,G^{2}\mu^{3}M^{2}r-11\,MGL^{2}\mu+18\,EL^{2}r\right )} \ .
%\nonumber\\
%&-& {\frac {8\,G^{3}L^{3}M^{2}}{45\,\mu\,r^{6}c^{7}}}
\end{eqnarray}
The momentary loss in $L_z$ is
\begin{eqnarray}
{dL_z\over dt} =
&+&{{8G^{2}LM\cos\iota\over 5\,\mu\,c^{5}r^{5}}\left (2\,E\mu\,r^{2}-3\,L^{2}\right )}\nonumber\\
&-& \frac {4\,G^{2}S}{15\,\mu^{2}c^{7}r^{7}} \{ [(6\,G^{2}M^{2}\mu
^{4}r^{2}+36\,\mu\,Er^{2}L^{2}+6\,GL^{2}M\mu^{2}r+18\,L^{4}+24\,\mu^{3
}Er^{3}GM )\sin^2\psi \nonumber\\&&
-39\,GL^{2}M\mu^{2}r+36\,L^{4}-36\,\mu
\,Er^{2}L^{2} ]\sin^2\iota+3\,r\dot r \, L\mu\,\sin2\psi\sin^2\iota
\left (2\,GM\mu^{2}r-3\,L^{2}+6\,E\mu\,r^{2}\right )\nonumber\\&&
-6\,G^{2}M^{2}\mu^{4}r^{2}+22\,GL^{2}M\mu^{2}r-24\,\mu^{3}Er
^{3}GM-36\,\mu\,Er^{2}L^{2} \} \ .
%-{\frac {8\,G^{3}L^{3}M^{2}\cos\iota}{45\,\mu\,r^{6}c^{7}}} \ .
\end{eqnarray}

  After parameterizing by $\chi$ and averaging over
the period $T$ as described in the previous section,
we obtain
\begin{eqnarray} \label{eq:avelosses}
\Biggl<{dE\over dt}\Biggr>=
&-&{4\,(-E\mu)^{3/2}G^{2}M\over 15\,c^{5}L^{7}\sqrt{2}} (148\,E^{2}L^{4}+732\,EG^{2}L^{2
}M^{2}\mu^{3}+425\,G^{4}M^{4}\mu^{6})\nonumber\\
% *** 1/c^7 terms neglected here: **************
%&-& {G^2(-2\,E)^{3/2}M\sqrt {\mu}\over 45\,c^{7}L^{6}}
%(12\,E^{2}L^{4}+60\,G^{2}EL^{2}M^{2}\mu^{3}+35\,G^{4}M^{4}\mu^{6})\\
%
&+& {2\,S\cos\iota G^2(-E\mu)^{3/2}\over 5\,c^{7}L^{10}\sqrt{2}}
(520\,E^{3}L^{6}+10740\,G^{2}E^{2}L^{4}M^{2}\mu^{3}
+24990\,G^{4}EL^{2}M^{4}\mu^{6}+12579\,G^{6}M^{6}\mu^{9})\nonumber\\
\Biggl<{dL\over dt}\Biggr>=
&-&{16\,G^2(-E\mu)^{3/2}M\over 5\,c^{5}L^{4}\sqrt {2}}
(14\,EL^{2}+15\,G^{2}M^{2}\mu^{3})\nonumber\\
% *** 1/c^7 terms neglected here: **************
%&-& {G^2(-2\,E)^{3/2}M\sqrt {\mu}\over 45\,c^{7}L^{6}}
%(12\,E^{2}L^{4}+60\,G^{2}EL^{2}M^{2}\mu^{3}+35\,G^{4}M^{4}\mu^{6})\\
%
&+& {4\,G^2(-E\mu)^{3/2}S\cos\iota \over 15\,c^{7}L^{7}\sqrt {2}}
(1188\,E^{2}L^{4}+6756\,G^{2}EL^{2}M^{2}\mu^{3}+5345\,G^{4}M^{4}\mu^{6})\nonumber\\
\Biggl<{dL_z\over dt}\Biggr>=
&-&\frac {16\,G^2(-E\mu)^{3/2}M\cos\iota}{5\,c^{5}L^{4}\sqrt {2}}(14\,EL^{2}+15\,G^{2}
M^{2}\mu^{3})\nonumber\\
% *** 1/c^7 terms neglected here: **************
%&-&\frac {G^2 (-2\,E)^{3/2}M\sqrt {\mu}\cos\iota}{45\,c^{7}L^{6}} (12\,E^{2}L^{4}+60\,G^{2}EL^{2}
%M^{2}\mu^{3}+35\,G^{4}M^{4}\mu^{6})\nonumber\\
%
&+& \frac {4G^{2}S(-E\mu)^{3/2}}{15\,c^{7}L^{7}\sqrt{2}} (1188\,E^{2}L^{4}
+6756\,G^{2}EL^{2}M^{2}\mu^{3}+5345\,G^{4}M^{4}\mu^{6})\\
&-&\frac {2G^{2}S(-E\mu)^{3/2}\sin^2\iota}{5\,c^{7}L^{7}\sqrt {2}}
(1172\,E^{2}L^{4}+5892\,G^{2}EL^{2}M^{2}\mu^{3}+4325\,G^{4}M^{4}\mu^{6})\nonumber\\
&+&\frac {8\,G^{2}S(-E\mu)^{3/2}\cos(2
\Psi_0)\sin^2\iota}{5\,c^{7}L^{7}\sqrt {2}}
(52\,E^{2}L^{4}+92\,G^{2}EL^{2}M^{2}\mu^{3}+33\,G^{4}M^{4}\mu^{6})
                                                  \ .\nonumber
\end{eqnarray}

    Among these losses averaged over one period of the radial
motion, the only quantity depending on the initial angle
$\Psi_0$ is $\bigl<{dL_z/ dt}\bigr>$. This is to be interpreted such
that the rate of loss of the angular momentum component $L_z$
varies with the position of the periastron. In (\ref{eq:avelosses}), the
terms with $\cos(2\Psi_0)$ average to zero when the precession
time scale is short compared to the radiation reaction time scale.
(For a more extensive discussion of this subject, {\em cf.} \cite{Ry2}.)
 
\section{Averaged Losses in Terms of Orbit Parameters}
 
 The major axis $a$ and the eccentricity $e$ of the Kepler motion
are constants and together with $\iota$ they determine the
full set of the
constants of motion $E,L$ and $L_z$. In the Lense-Thirring picture of a
perturbed Kepler motion $E,L$ and $L_z$ contain first-order
terms, and the corresponding constants $a$ and $e$ should also contain
first-order terms. There is an ambiguity in how to introduce
these orbit parameters, as noted by Ryan\cite{Ry2}. He chooses
these parameters by referring to the
quasicircular orbits in the Kerr metric. This method is passing
when wishing to approximate Carter orbits. However,
there appears to be no reason for invoking phenomena tied to
black holes when describing arbitrary weakly bound spinning
bodies. In fact, we want to keep our treatment general enough to
cover all axially symmetric gravitational fields with first order
contributions from the spin-orbit interaction term
in the Lagrangian (\ref{delag}). Thus we choose as the definition
of the semimajor axis $a$ and eccentricity $e$:
\begin{equation}
r_{{}_{min}^{max}}=a(1\pm e)\ ,
\end{equation}
which implies:
 
\begin{eqnarray}
E&=&-{GM\mu\over 2a}\left(1+{2GS\cos\iota\over c^2a^{3\over 2}
\sqrt{GM(1-e^2)}}\right)\\
L^2&=&GM\mu^2a(1-e^2)\left(1-{2GS\cos\iota\over c^2a^{3\over 2}
\sqrt{GM(1-e^2)}}{3+e^2\over 1-e^2}\right)\ .
\end{eqnarray}
These relations can be inverted:
\begin{eqnarray}
\label{eq:aa}
a&=&-{GM\mu\over 2E}\left(1-{4EL_zS\over
c^2ML^2}\right)\label{aa}\\
\label{eq:ee}
1-e^2&=&-{2EL^2\over G^2M^2\mu^3}\left[1+{8L_zS\over c^2ML^4}
(G^2M^2\mu^3+EL^2)\right]\ .\label{ee}\
\end{eqnarray}
 
The averaged losses in terms of $a$ and $e$ then read:
\begin{eqnarray}
\Biggl<{dE\over dt}\Biggr>=
&-&(37\,e^{4}+292\,e^{2}+96)\frac {G^{4}M^{3}\mu^{2}}{15\,(1-e^2)^{7/2}a^{5}c^{5}}
%-(9\,e^{6}+138\,e^{4}+152\,e^{2}+16)\frac {G^{5}M^{4}\mu^{2}}
%{90\,(1-e^2)^{9/2}a^{6}c^{7}}
\nonumber\\
&+&(491\,e^{6}+5694\,e^{4}+6584\,e^{2}+1168)
\frac{G^{9/2}SM^{5/2}\mu^{2}\cos\iota}{30\,c^7a^{13/2}(1-e^2)^{5}}\\
\Biggl<{dL\over dt}\Biggr>=
&-&(7\,e^{2}+8)\frac{ 4G^{7/2}M^{5/2}\mu^{2}}{5\,a^{7/2}
(1-e^2)^{2}c^{5}}
%-(3\,e^{4}+24\,e^{2}+8)
%\frac{G^{9/2}M^{7/2}\mu^{2}}{45\,c^7a^{9/2}(1-e^2)^{3}}\nonumber\\
 + (549\,e^{4}+1428\,e^{2}+488)
\frac{G^{4}SM^{2}\mu^{2}\cos\iota}{15\,c^{7}a^{5}(1-e^2)^{7/2}}\\
\Biggl<{dL_z\over dt}\Biggr>=
&-&(7\,e^{2}+8)\frac{4G^{7/2}M^{5/2}\mu^{2}\cos\iota}{5\,c^5a^{7/2}(1-e^2)^{2}}
%-(3\,e^{4}+24\,e^{2}+8)\frac{G^{9/2}M^{7/2}\mu^{2}\cos\iota}{
%45\,c^7a^{9/2}(1-e^2)^{3}}
\nonumber\\
&-&(285\,e^{4}+1512\,e^{2}+488)\frac{G^{4}SM^{2}\mu^{2}}{
30\,c^{7}a^{5}(1-e^2)^{7/2}}+(461\,e^{4}+1456\,e^{2}+488)\frac{
G^{4}SM^{2}\mu^{2}\cos^2\iota}{10\,c^{7}a^{5}(1-e^2)^{7/2}}\nonumber\\
&+&(13\,e^{2}+20)2e^{2}\frac{G^{4}SM^{2}\mu^{2}\sin^2\iota\cos(2 \Psi_0)}{
5\,c^{7}a^{5}(1-e^2)^{7/2}} \ . \nonumber\\
\end{eqnarray}
 
Differentiating (\ref{iota}) one has:
\begin{equation}
\label{eq:aviota}
\Biggl<{d\iota\over dt}\Biggr>=
   {1\over L \sin\iota}\Biggl[\cos\iota\Biggl<{dL\over dt}\Biggr>
                          -\Biggl<{dL_z\over dt}\Biggr>\Biggr]\ .
\end{equation}
   Those terms in the brackets which are zero-order in $S$, cancel.
 
 The averaged losses of the ellipse parameters can be obtained by
differentiating (\ref{aa}) and (\ref{ee}) and from (\ref{eq:aviota}):
 \begin{eqnarray}
\Biggl<{de\over dt}\Biggr>&=&
-\frac{G^3M^{2}\mu e}{15\,(1-e^2)^{5/2}a^{4}c^{5}}
\Biggl({121} e^2 + {304}\Biggr)
+\frac{G^{7/2}M^{3/2}e\mu}{(1-e^2)^4 a^{11/2}c^7}S
\cos\iota\Biggl(\frac{1313}{30} e^4 + \frac{932}{5}e^2 +
\frac{1172}{5}\Biggr)
         \\     \label{eq:dadt}
\Biggl<{da\over dt}\Biggr>&=&
-\frac{64G^3 M^2\mu}{5a^3c^5(1-e^2)^{7/2}}
          \Biggl(\frac{37}{96} e^4 + \frac{73}{ 24} e^2 + 1\Biggr)
+\frac{64}{5}\frac{G^{7/2}M^{3/2}\mu S\cos\iota}{a^{9/2}c^7(1-e^2)^{5}}
\Biggl( \frac{121}{ 64} e^6 + \frac{585}{ 32}
e^4 + \frac{124}{ 3} e^2 + \frac{133}{ 12}\Biggr)
         \\     \label{eq:didt}
\Biggl<{d\iota\over dt}\Biggr>&=&
S\sin\iota
\frac{G^{7/2}M^{3/2}\mu }{a^{11/2}c^7 (1-e^2)^4}
  \Biggl[ - \cos (2\Psi_0)\Biggl(\frac{26}{ 5} e^4 +
8 e^2\Biggr) + \frac{19}{ 2} e^4 + \frac{252}{ 5} e^2 +
\frac{244}{15}\Biggr]           \ .
\end{eqnarray}

Let us take an example of a binary system contaning a massive black
hole with mass $100M_\odot$ and spin $S\approx 0.6M^2$, and
where the companion's orbit has the semimajor axis $10^4$ km,
{\em i.e.}, one-hundredth of the Hulse-Taylor system.
Then the increase in the angle $\iota$ per loss of $\log a$
(a dimensionless quantity) is
$<d\iota/d\log a>=(G/m/a^3)^{1/2}S \sin\iota\approx 10^{-3}\sin\iota$
for a circular orbit.
 
The enhancement factor for an eccentric orbit is
\begin{equation}
f(e)=
{ 285 e^4 + 1512 e^2 + 488 - 6 (26 e^4 +40 e^2 )\cos(2\Psi_0)
\over 4 (37 e^4 + 292 e^2 + 96 )(1-e^2)^{1/2} }\ .
\end{equation}
 
\begin{figure}[htb]
\hspace*{1.6in}
\special{hscale=35 vscale=35 hoffset=-20.0 voffset=20.0
         angle=-90.0 psfile=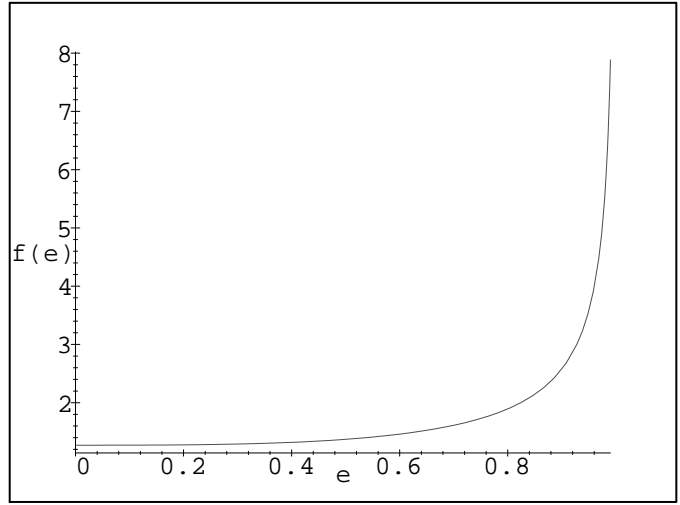}
\vspace*{2.6in}
%\psfig{psfile=num01.eps,height=4cm,width=6cm}
%\epsfysize=8cm
%\centerline{\hfill
%\epsfbox{num01.eps}\hfill}
\caption{The enhancement factor.}
\end{figure}
 
  The enhancement factor is monotonously increasing (Fig.1) from
$f(0)=122/121$ to make the change in $\iota$ comparable to that in $\log
a$ only when the eccentricity approaches the unit value to one part in
$10^6$.
 
\section{Concluding Remarks}
 
The Kepler orbits are completely characterized by the energy $E$,
the angular momentum vector ${\bf L}$ and the Runge-Lenz vector
${\bf A}$ at the initial time $t_0$. These are six independent
constants of the motion. In the Lense-Thirring
picture, the constants of the motion alone do not suffice for the full
description of the orbit. We complete the characterization by specifying
the Euler angles $\Psi_0$ and $\Phi_0$ of the periastron. Their
evolution over one period,
under the influence of the spin (disregarding radiation backreaction)
is given in (\ref{dephi}) and (\ref{depsi}).
Taking into account the gravitational radiation backreaction, the shifts
(\ref{depsi}) and (\ref{dephi}) will vary from one period to another, as
the 'constants' $L$ and $L_z$ evolve. One may interpret these as the
radiation-induced changes in the parameters $\Psi_0$ and $\Phi_0$:
\begin{equation}
\label{eq:dpsi}
\delta\Psi_0\ :=\ \Delta\Psi (T)-\Delta\Psi (0)\ =\
2\pi S{18G^2M\mu^3\over c^2L^4}\delta L_z  \\
\end{equation}
\begin{equation}
\label{eq:dphi}
\delta\Phi_0\ :=\ \Delta\Phi (T)-\Delta\Phi (0)\ =\
-2\pi S{6G^2M\mu^3\over c^2L^4}\delta L  \ ,\\
\end{equation}
where $\delta L=T \bigl<{dL/dt}\bigr>$ and we have used
that in zeroth order $\delta L_z=\delta L \cos\iota$.
The total shift in the Euler angle $\Psi_0$ of the periastron point
is given by the sum of (\ref{eq:depsi}) and (\ref{eq:dpsi}), and
likewise for $\Phi_0$.
 
We continue with some comments on the radiative losses in the constants
of motion. The terms independent of $S$ agree with the results of
Peters and Mathews\cite{PM,P}.
The terms linear in the spin $S$ in our averaged losses agree with
Ryan's corresponding results, when the parameters $a$, $e$ and
$\iota$ are converted to his parameters. In the present paper, the
energy and angular momentum losses have been directly computed.
This is to be contrasted with Ryan's method where the backreaction
forces have been used. The perfect agreement achieved underscores
the reliability of these computations.
 
When $\iota=0$, the averaged losses in $L$ are equal to the
averaged losses in $L_z$ up to first order in $S$. Hence the
equatorial plane of the orbits is stable with respect to radiation
losses.
 
  As we have shown in the previous section, the inclination
of the orbit cannot change significantly in the adiabatic era.
Hence we conclude that it is
insufficient to restrict the computation of signal
templates from a black hole -- test particle system
to equatorial Carter orbits.
 
In the limiting case of circular orbits, $e\to0$, there is no
distinguished point on the ellipse. The dependence of
$\bigl<{dL_z/dt}\bigr>$ on $\Psi_0$ vanishes.
As expected from the axial symmetry, the
radiative losses, for arbitrary excentricity, do not depend on the
initial angle $\Phi_0$.
 
  In the present paper, we have developed a toolchest for the
computation of radiation losses, which include a new
parametrization of the orbit, new angular variables and the
application of the residue
theorem for obtaining time averages. Currently we are employing
these procedures with the inclusion of finite mass
effects.\cite{GPV}

\section{Acknowledgments}
 
  This work has been supported by OTKA no. T17176 and D23744
grants. The algebraic packages REDUCE and MapleV have been used for
checking our computations.
 
\appendix
\section{Alternative Angular Variables}
 
Introduce three angle variables $\Psi^\prime, \iota^\prime$ and
$\Phi^\prime$, where $\Psi^\prime$ is the angle of rotation about
${\bf L}_N$, bringing the position of the particle to the node;
$\iota^\prime$ is the angle subtended by
${\bf L}_N$ and ${\bf S}$ and $\Phi^\prime$ is the angle of
precession of ${\bf L}_N$ about ${\bf S}$.
The angle $\Psi^\prime$ is the {\em argument of the latitude},
$\iota^\prime$ the {\em inclination} and $\Phi^\prime$ the
{\em longitude of the node} \cite{LT}.
Unlike $\iota$, the angle $\iota^\prime$ is not constant.
Let us express $\iota^\prime$ in terms of $\iota$ and $\Psi$.
Computing the $z$ component of ${\bf L}_N$ in two different ways
we get the two sides of the equation
\begin{equation}
\{\ \mu^2r^4\ (\dot\theta^2+sin^2\theta\ \dot\varphi^2)\ \}
^{1\over 2}\
   \cos\iota^\prime=\mu r^2\sin^2\theta\ \dot\varphi \ .
\end{equation}
With the time derivatives of polar angles (\ref{polar}) after
linearization in $S$, the previous equation gives
 
\begin{equation}
 \cos\iota^\prime=\cos\iota\bigl(1+{2G\mu S\over c^2L_zr}\
   \sin^2\iota\cos^2\Psi\bigr) \ .
\end{equation}
Here we are using that the primed angles are equal, to zeroth order in
$S$, with the corresponding unprimed ones. Taking the time derivative
of $\iota^\prime$ we get
\begin{equation}
   \dot\iota^\prime={2GS\sin\iota\cos\Psi \over c^2Lr^3}
   (\mu r\dot r\cos\Psi+2L\sin\Psi) \ .
\end{equation}
With the new time-dependent Euler rotations the coordinates
of the particle become
\begin{eqnarray}
x&=&r(\cos\Phi^\prime\cos\Psi^\prime-\cos\iota^\prime\sin\Phi^\prime\sin\Psi^\prime) \nonumber\\
y&=&r(\sin\Phi^\prime\cos\Psi^\prime+\cos\iota^\prime\cos\Phi^\prime\sin\Psi^\prime) \\
z&=&r \sin\iota^\prime\sin\Psi^\prime \nonumber\
\end{eqnarray}
and the time derivatives of the polar angles are
\begin{equation}
   \dot\theta=-{\sin\iota^\prime\cos\Psi^\prime\over \sin\theta}\ \dot\Psi^\prime-
   {\cos\iota^\prime\sin\Psi^\prime\over \sin\theta}\ \dot\iota^\prime\ ,\qquad
   \sin^2\theta=1-\sin^2\iota^\prime \sin^2\Psi^\prime\ ,
\end{equation}
\begin{equation}
   \dot\varphi=\dot\Phi^\prime+
    {\cos\iota^\prime\over \sin^2\theta}\ \dot\Psi^\prime-
    {\sin\iota^\prime\sin 2\Psi^\prime\over 2\sin^2\theta}\ \dot\iota^\prime
     \ .
\end{equation}
Inserting these equations in (\ref{polar}), we get the changes of
the primed Euler angles in terms of the unprimed ones
\begin{equation}
   \dot\Psi^\prime={L\over \mu r^2}+{GS\cos\iota\over c^2Lr^3}\
   (2L-\mu r\dot r\sin 2\Psi)\
\end{equation}
\begin{equation}
   \dot\Phi^\prime={2GS\sin\Psi\over c^2Lr^3}\
   (\mu r\dot r\cos\Psi+2L\sin\Psi)\ .
\end{equation}
Integrating these equations, we get the new angle variables and their
shifts over one period
\begin{equation}
   \Delta\iota^\prime= 0 ,\qquad\
   \Delta\Psi^\prime= 2\pi\ -\ 2\pi S\frac{6G^2M\mu^3L_z}{c^2L^4}
     , \qquad
   \Delta\Phi^\prime= 2\pi S\frac{2G^2M\mu^3}{c^2L^3} \ .
\end{equation}
Note that these changes in the primed variables are equal to the
shifts (\ref{dephi}) and (\ref{depsi}) of the unprimed variables.

\end{document}